\documentclass[aps, prb, twocolumn, superscriptaddress]{revtex4-2}

\usepackage{cmap}

\usepackage[english]{babel}
\usepackage{amsmath}
\usepackage{bm}

\usepackage{graphicx}
\usepackage[caption=false]{subfig}

\usepackage{slashbox}
\usepackage{tabularx}

\usepackage[bookmarks=false]{hyperref}
\hypersetup{
    colorlinks=true,
    linkcolor=blue,
    citecolor=blue,
    urlcolor=blue
}

\begin{document}

\title{Hydrogen modes in KH$_2$PO$_4$ under pressure from \textbf{\textit{ab initio}} calculation \\ and inelastic neutron scattering}

\author{V. A. Abalmasov}
\email{v.a.abalmasov@math.nsc.ru}
\affiliation{Sobolev Institute of Mathematics, 630090 Novosibirsk, Russia}
\author{A. S. Ivanov}
\affiliation{Institut Max von Laue - Paul Langevin, 38042 Grenoble, France}
\author{R. A. Sadykov}
\affiliation{Institute for Nuclear Research RAS, 117312 Moscow, Russia}
\author{A. V. Belushkin}
\affiliation{International Intergovernmental Organization Joint Institute for Nuclear Research, Frank Laboratory of Neutron Physics, 141980 Dubna, Russia}
\affiliation{Institute of Physics, Kazan Federal University, 420008 Kazan, Russia}

\date{\today}

\begin{abstract}

The nature of the phonon triplet in the region of OH-stretching modes in hydrogen-bonded materials is often explained by the interplay of OH-stretching modes and combinations and overtones of OH-bending modes. In order to elucidate the both contributions in KH$_2$PO$_4$ (KDP), we compare the pressure dependence of the OH-bending and stretching modes from {\it ab initio} calculation and inelastic neutron scattering (INS) measurements. The {\it ab initio} calculation predicts a hardening of OH-bending modes and a softening of OH-stretching modes with pressure. At the same time, INS measurements in the region of OH-stretching modes indicate a hardening of the phonon triplet together with the bending modes. This means that this triplet in INS measurements is mainly due to combinations and overtones of OH-bending modes, while the intensity of OH-stretching modes appears to be relatively low. This conclusion may also apply to other hydrogen-bonded materials.

\end{abstract}

\maketitle

\section{Introduction}

Potassium dihydrogen phosphate (KH$_2$PO$_4$, or KDP) is an archetypal order-disorder type ferroelectric~\cite{lines2001}. Due to its unique nonlinear optical properties, combined with a high optical damage threshold and reproducible growth to large sizes~\cite{zaitseva2001, zaitseva2022}, it has found wide application as electro-optic switches and frequency converters in high-power laser systems~\cite{mironov2022}, and is currently irreplaceable in inertial confinement fusion engineering~\cite{hawley-fedder2004}. KDP is also the first ferroelectric to be identified with bistable hydrogen \mbox{OH--O} bonds~\cite{busch1935}, widespread in nature from the large family of H-bonded ferroelectrics~\cite{horiuchi2010} and water to complex DNA molecules~\cite{jeffrey1991, marechal2006, grabowski2006}. H-bonds also play a crucial role in ferroelectric polarization switching in KDP~\cite{fancher2021} and may be harnessed in the future for all-optical polarization switching~\cite{larsson2004, zamponi2012, mankowsky2017, abalmasov2020, abalmasov2021b, gonzalez-vallejo2022}. Nevertheless, despite a long history of research, the triplet phonon structure in the region of OH-stretching modes in KDP is still not fully understood.

\begin{figure}[]
\centering
\vspace{0.3cm}
\includegraphics[width= 0.9\columnwidth]{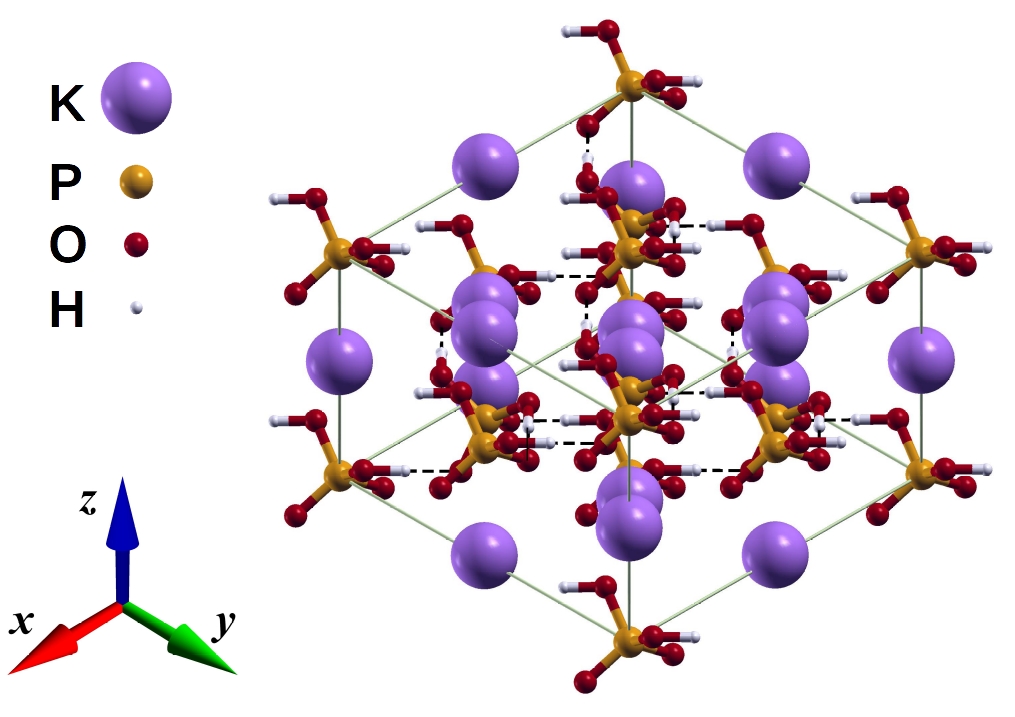}
\vspace{-0.cm}
\caption{KDP conventional unit cell in the ferroelectric phase. Hydrogen bonds (shown as black dashed lines) lie in the $xy$ planes, connecting two oxygen atoms in adjacent phosphate tetrahedra.}
\label{fig:1}
\end{figure}

The crystal structure of KDP below the ferroelectric phase transition at $T_c = 122$~K is orthorhombic with the space group $Fdd2$ ($C^{19}_{2v}$)~\cite{nelmes1987}. The primitive unit cell contains two formula units and the conventional unit cell is face-centered with eight formula units~(Fig.~\ref{fig:1}). In deuterated KDP (DKDP), the phase transition temperature is much higher, $T_c = 229$~K, which indicates the decisive role of hydrogen in the ferroelectic phase transition~\cite{nelmes1987}. Indeed, in the paraelectric (PE) phase, hydrogen atoms are disordered in a double-well potential along the hydrogen bonds lying in the {\it xy} plane, while being ordered in the ferroelectric (FE) phase.

Since hydrogen is the lightest ion and strongly bonded to oxygen, the fundamental OH vibrational frequencies are the highest in the phonon spectrum of the crystal. The OH-modes are further distinguished by their softening upon deuteration by about $(m_{\text{OD}}/m_{\text{OH}})^{1/2} \approx 1.37$ in accordance with the increase in the square root of the oxygen-hydrogen reduced mass $m_{\text{OH}}$ and the assumption of a harmonic potential~\cite{tominaga2003}. The phonon spectrum of KDP~(DKDP) was obtained by various methods, including inelastic neutron scattering (INS)~\cite{shibata1992, mizoguchi1993, belushkin1993JPSJ, belushkin1993ZPB, ikeda1995, belushkin1997}, infrared (IR) measurements~\cite{blinc1958, imry1965, hill1968, coignac1971, cody1975, khanna1980, gervais1987, simon1988}, Raman scattering~\cite{she1971, she1972, tominaga2003, mita2006, liu2007, malinovsky2008, malinovsky2009, abalmassov2011, abalmassov2018, takenaka1992}, impulsive stimulated Raman scattering~\cite{dougherty1992, yoshioka1998, yoshioka1998b, yoshioka1999, yagi2000, kano2002, kano2003, kano2004, kano2006}. The frequencies of the OH-modes were identified as 1012, 1312, 1856, 2400, and 2680~cm$^{-1}$~\cite{tominaga2003} (we note that in INS measurements, phonon energies are usually presented in units of meV, and at 1~cm$^{-1} \approx 0.124$~meV these bands correspond to 125, 163, 230, 298, 332~meV). The two lower bands are narrow and well resolved, but are masked in the IR and Raman measurements by the intense phosphate tetrahedra internal stretching modes in the $800 - 1000$~cm$^{-1}$ region~\cite{tominaga2003, mita2006, liu2007, tominaga2010, abalmassov2018}. In the broad phonon structure in the region of $1500 - 3000$~cm$^{-1}$ ($190 - 370$~meV), a triplet of three upper wide bands is clearly visible. Their widths are about $200 - 300$~cm$^{-1}$ ($25 - 40$~meV) in the PE phase with further broadening and slight splitting in the FE phase~\cite{abalmassov2018}. However, the assignment of hydrogen modes in KDP is still controversial.

There is a general consensus that the bands of 1000 and 1300~cm$^{-1}$ (125 and 160~meV) correspond to the out-of-plane $\gamma_{\text{OH}}$ and in-plane $\beta_{\text{OH}}$ (sometimes referred to as $\delta_{\text{OH}}$ in the literature~\cite{novak1974}) bending vibrations, respectively~\cite{liu2007}. In turn, the position and structure of the OH-stretching vibrations $\nu_{\text{OH}}$ are strongly affected by the anharmonicity of the H-bond potential, its dependence on the bond length, and the possible Fermi resonance with the overtones and combinations (i.e. with the emission of a pair of phonons belonging to the same or different modes, respectively) of $\gamma_{\text{OH}}$, $\beta_{\text{OH}}$ and PO$_4$-stretching modes, which can acquire greater intensity as a result~\cite{bratos1991}. Together, this is supposed to determine the triplet structure often observed in materials with strong hydrogen bonds as calculated for some model parameters in~\cite{bratos1982, bratos1991}.

On the other hand, it was noted that the hydrogen bond length (i.e. the distance between the bound oxygen atoms) of about $R_{\text{OO}}=2.49$~\AA\, in KDP under ambient conditions~\cite{nelmes1987} corresponds to the stretching vibration frequency of about 1300~cm$^{-1}$ (160~meV) according to the statistical correlation diagram for hydrogen-bonded materials~\cite{novak1974, bratos1991}. This was the argument in~\cite{mita2006} to assign this frequency to $\nu_{\text{OH}}$. The same interpretation was adopted in~\cite{shibata1992}, where the INS band at 159~meV (1282~cm$^{-1}$) was assigned to $\beta_{\text{OH}}$ and 161~meV (1299~cm$^{-1}$) to $\nu_{\text{OH}}$. However, we note that the KDP covalent OH-bond length $r_{\text{OH}}=1.07$~\AA\, under ambient conditions~\cite{nelmes1987} and its $\gamma_{\text{OH}}$ mode frequency of 1000~cm$^{-1}$ (125~meV) both correspond to the stretching vibration frequency of about 2700~cm$^{-1}$ (330~meV) according to the corresponding correlation diagrams~\cite{novak1974}. As mentioned in~\cite{bratos1991}, the correlation diagrams are not exact, especially for strong bonds. At the same time, according to the results of neutron Compton scattering experiments, the stretching vibration frequencies of 2039~cm$^{-1}$ (253~meV) in KDP~\cite{reiter2002} and only 951~cm$^{-1}$ (118~meV) in DKDP~\cite{reiter2008} were claimed. The authors explain the large difference in these two frequencies by the anharmonicity of the potential along the hydrogen bonds and its dependence on deuteration.

Although {\it ab-initio} calculations in KDP are quite numerous~\cite{hao1992, zhang2001, koval2002, koval2005, lasave2005, lasave2009, srinivasan2011, wikfeldt2014, jia2020, jiaM2020, yang2021}, they mostly consider its structure and energy, and only a few recently published works are devoted to its phonon spectrum~\cite{menchon2018, engel2018, jiang2020}. However, the later diverge in their prediction of the OH-stretching frequency: it is about 2700~cm$^{-1}$ (330~meV) as calculated in~\cite{menchon2018} and $2250 - 2400$~cm$^{-1}$ ($280 - 300$~meV) according to~\cite{engel2018, jiang2020}, which also leaves ambiguity.

\begin{figure}[t]
\centering
\includegraphics[width= 0.9\columnwidth]{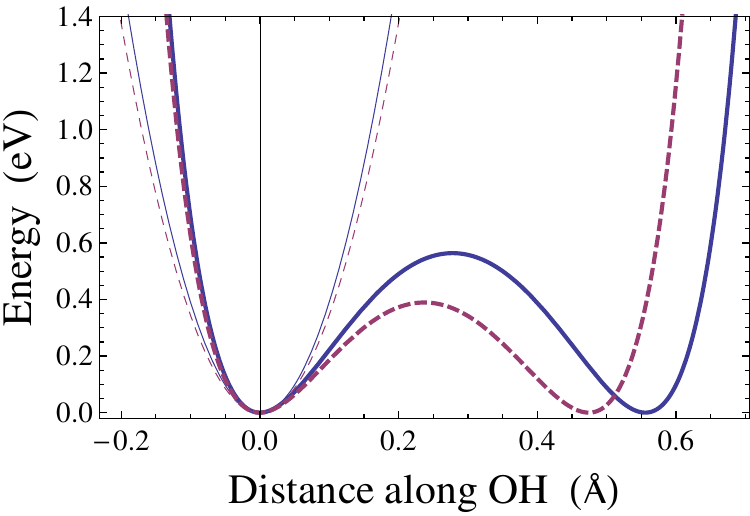}
\vspace{-0.cm}
\caption{Potential energy surface of a hydrogen atom along an H-bond approximated by the double-Morse potential~\cite{lawrence1980, lawrence1981, robertson1981} for larger (thick solid blue line) and shorter (thick dashed purple line) distances between bonded oxygen atoms. The thin lines correspond to the harmonic approximation of the double-Morse potential in the vicinity of one of its two minima. The harmonic potential frequency decreases with the distance between the two oxygen atoms.}
\label{fig:energy_pressure}
\end{figure}

The length of the hydrogen bond $R_{\text{OO}}$ turns out to be the key parameter determining the frequency of the OH-stretching mode~\cite{novak1974}. Moreover, the bending and stretching vibration frequencies should depend differently on $R_{\text{OO}}$. Indeed, as the interatomic distances decreases, the bending vibration frequencies in a single-well potential in the plane perpendicular to the H-bond increases. At the same time, the stretching vibration frequency in the double-well potential along the bond, which is determined by the second derivative of the potential at its minimum, should decrease until the barrier disappears (see Fig.~\ref{fig:energy_pressure}, where the double-Morse potential for the H-bond in KDP is used for illustration~\cite{lawrence1980, lawrence1981, robertson1981}). This is consistent with the bending-stretching vibration frequency correlation diagrams~\cite{novak1974}. Thus, by controlling the distance, we could clearly distinguish between bending and stretching vibration modes according to their behavior. In turn, $R_{\text{OO}}$ increases with temperature and decreases with pressure. Note that the softening in the OH-stretching vibrations under pressure was observed by Raman and IR spectroscopy in water ice Ih~\cite{minceva1984}, VI~\cite{abebe1979, walrafen1982}, VII~\cite{zha2016, song1999}, VIII~\cite{pruzan2003}, and liquid water~\cite{kawamoto2004, li2020}. The hardening of the OH-stretching mode in ice Ih with increasing temperature was observed, e.g., in \cite{wang2019, fukazawa2000} using the same techniques and in~\cite{jaiswal2023} by INS (in the later case, the bending mode hardens with temperature as well). However, a hardening of the OH-stretching modes for water ice I$h$ with pressure and for higher pressure phases of water ice is observed in INS measurements in~\cite{straessle2004, li1992, li1996}, which also contradicts the results of {\it ab-initio} calculations~\cite{tian2012}.

The temperature dependence of the hydrogen modes in KDP has recently been studied using Raman spectroscopy~\cite{abalmassov2018}. It was observed that the 1300 and 1800~cm$^{-1}$ (160 and 220~meV) modes hardened upon cooling in the PE phase, while the positions of the modes 2400 and 2700~cm$^{-1}$ (300 and 330~meV) did not change much, perhaps with a very slight tendency to softening, which may correspond to a decrease in the barrier potential of the double-well hydrogen bond as discussed above (see Fig.~\ref{fig:energy_pressure}). As noted in~\cite{robertson1981}, the changes in atomic positions in KDP that occur on cooling from 295 to 128~K, i.e. in the PE phase, can be obtained by applying a hydrostatic pressure of only 0.6~GPa. Thus, the experimentally accessible pressure range may allow more clear observation of changes in the frequencies of OH-modes.

So far, only the 1000 and 1300~cm$^{-1}$ (125 and 160~meV) modes have been investigated under pressure using Raman spectroscopy~\cite{mita2006}, which showed their hardening, characteristic of the single-well bending mode potential. INS at a pressure of 2~GPa and a temperature of 30~K in the energy range up to 300~meV~\cite{belushkin1997} indicated a significant hardening of the $\gamma_{\text{OH}}$ and broadening of the $\beta_{\text{OH}}$ bending modes, and disappearance of modes at 238~meV (1920~cm$^{-1}$) and about 270~meV~(2180~cm$^{-1}$).

\begin{figure}[]
\centering
\includegraphics[width= 0.9\columnwidth]{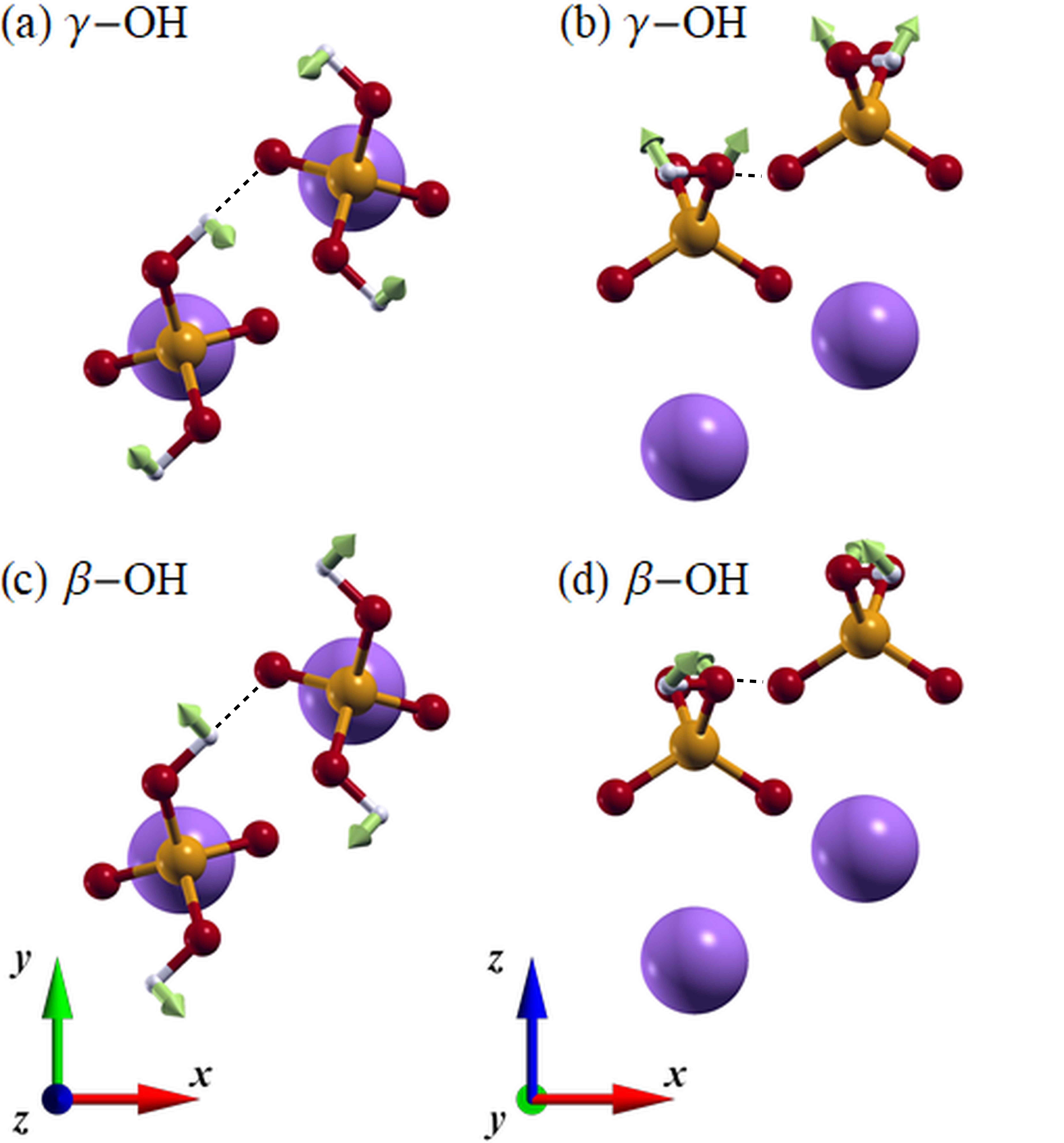}
\vspace{-0.cm}
\caption{Calculated {\it ab initio}  out-of-plane $\gamma_{\text{OH}}$ and  in-plane $\beta_{\text{OH}}$ OH-bending modes with symmetry $A_1$. Arrows correspond to ion displacements. }
\label{fig:bending}
\end{figure}

In the present work, we perform {\it ab initio} calculation of the OH-mode phonon spectrum of KDP and carry out INS measurements up to 350 meV, i.e. including the highest OH-mode fundamental frequency, and pressure up to 2.5~GPa, and compare the results. INS provides a unique method to study the vibrational dynamics of hydrogen atoms due to the large incoherent neutron scattering cross section of hydrogen. With this comparison we want to reveal the nature of the triplet in the phonon spectrum in the range of OH-stretching modes and try to assign the OH-modes to the measured spectral lines. In what follows, Sec.~\ref{abinitio} is devoted to our {\it ab initio} calculation of the structural parameters and optical phonon modes in KDP at various pressures. The results of our INS measurements are presented in Sec.~\ref{INS}.  In Sec.~\ref{discussion}, we compare and discuss the results of both approaches. We draw conclusions in Sec.~\ref{conclusions}.

\section{{Ab initio} calculation}\label{abinitio}

{\it Ab initio} calculation was performed using QUANTUM ESPRESSO~(QE) package \cite{giannozzi2009, giannozzi2017}. The generalized gradient approximation (GGA) of Perdew {\it et al.} (PBE)~\cite{perdew1996} was employed as an exchange-correlation potential. The valence electrons were represented by projector augmented wave (PAW) pseudopotentials, and the Brillouin zone (BZ) integral sampling $4\times 4\times 4$ was expressed with Monkhorst-Pack scheme k-points samplings~\cite{monkhorst1976}, which proved sufficient to achieve converged results. The cutoff energy of structural optimization was 900~eV. The phonon spectrum is calculated within the framework of density functional perturbation theory~\cite{baroni2001}. The phonon density of states (DOS) was calculated using a uniform $3\times 3\times 3$ q-point grid.

\begin{figure}[t]
\centering
\includegraphics[width= 0.9\columnwidth]{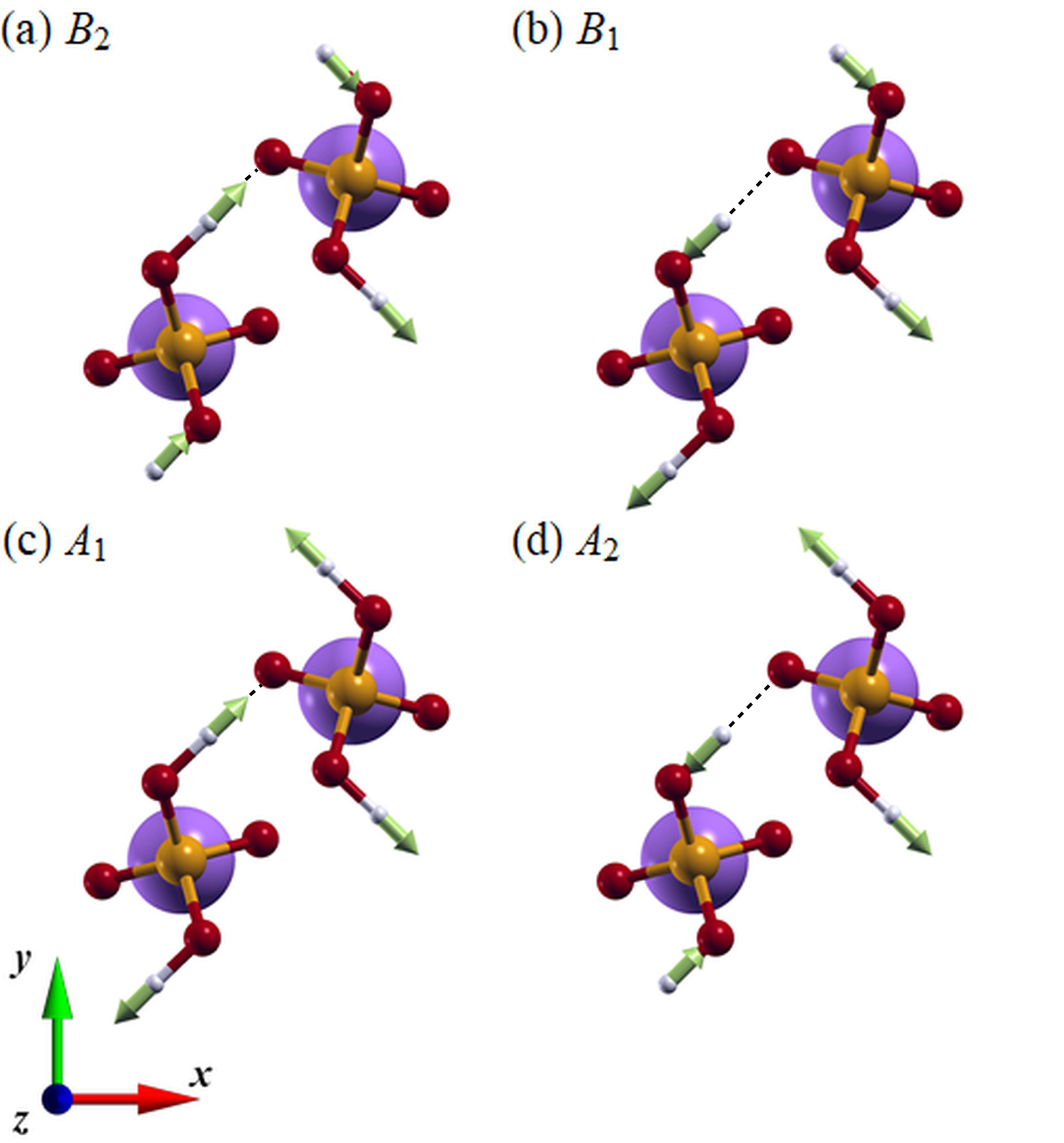}
\vspace{-0.cm}
\caption{Calculated {\it ab initio} OH-stretching modes $\nu_{\text{OH}}$ according to their irreducible representations. Arrows correspond to ion displacements.}
\label{fig:stretching}
\end{figure}

\begin{table}[b]
\setlength{\tabcolsep}{3pt}
\caption{Calculated {\it ab initio} frequencies of transverse optical (TO) modes in KDP (in ascending order towards the bottom of the table) at different pressures $p$ at the $\Gamma$-point of the Brillouin zone. The last column gives the frequencies of IR-active longitudinal optical (LO) modes $A_1(z)$, $B_1(y)$, $B_2(x)$ with their polarization axis indicated in parentheses. $A_2$ mode is not IR-active and does not experience LO-TO splitting. All modes are Raman active \cite{simon1988}.}
\label{table:freq}

\centering
\resizebox{\columnwidth}{!}{%
\begin{tabular}{ c | c @{\hspace{-0pt}} c @{\hspace{-0pt}} c | c  }
\hline 
\hline
Irrep & & $\nu_{\text{TO}}$, cm$^{-1}$ (meV) & & $\nu_{\text{LO}}$ \\ 
\cline{2-5}

 & $p = 0$ & $p = 1.4$~GPa & $p = 2.5$~GPa  & $p = 0$   \\ \hline

\multicolumn{5}{c}{$\gamma_{\text{OH}}$ out-of-plane bending} \\

$A_2$ & 1061 (131.6) & 1080 (133.9) & 1093 (135.5) & 1061 (131.6)\\   
$A_1$ & 1067 (132.3) & 1091 (135.3) & 1107 (137.3) & 1107 (137.3)\\  
$B_2$ & 1081 (134.0) & 1107 (137.3) & 1124 (139.4) & 1081 (134.0)\\   
$B_1$ & 1086 (134.7) & 1110 (137.6) & 1127 (139.7) & 1096 (135.9)\\  \hline

\multicolumn{5}{c}{$\beta_{\text{OH}}$ in-plane bending} \\

$A_1$ & 1285 (159.3) & 1296 (150.7) & 1303 (161.6) & 1295 (161.6)\\   
$A_2$ & 1288 (159.7) & 1297 (160.8)& 1304 (161.7)& 1288 (159.7)\\  
$B_2$ & 1304 (161.7)& 1309 (162.3)& 1312 (162.7)& 1324 (164.2)\\  
$B_1$ & 1307 (162.1)& 1319 (163.6)& 1327 (164.5)& 1318 (163.4)\\ \hline

\multicolumn{5}{c}{$\nu_{\text{OH}}$ stretching} \\

$B_2$ & 2232 (276.8)& 2125 (263.5)& 2051 (254.3)& 2428 (301.1)\\   
$B_1$ & 2234 (277.0)& 2128 (263.9)& 2054 (254.7)& 2439 (302.4)\\   
$A_1$ & 2237 (277.4)& 2131 (264.2)& 2056 (254.9)& 2253 (279.4)\\  
$A_2$ & 2414 (299.3)& 2317 (287.3)& 2250 (279.0)& 2414 (299.3)\\   
\hline 
\hline
\end{tabular}
}
\end{table}

The frequencies of four ``out-of-plane'' OH-bending modes $\gamma_{\text{OH}}$ at $\Gamma$-point and zero pressure are from 1061 up to 1086~cm$^{-1}$ (see Table~\ref{table:freq}) and are about 50~cm$^{-1}$ higher than their experimental value of 1012~cm$^{-1}$~\cite{tominaga2003}. Their strong dependence on pressure (with an average pressure coefficient of 15.7~cm$^{-1}$GPa$^{-1}$ $\approx 1.95$~meV GPa$^{-1}$) is in accordance with the present INS measurements (see below) and Raman scattering data~\cite{mita2006}, where a pressure coefficient of 20.6~cm$^{-1}$ GPa$^{-1}$ was obtained. The calculated frequencies of the ``in-plane'' OH-bending modes $\beta_{\text{OH}}$ at $\Gamma$-point and zero pressure are very close to their experimental value of 1312~cm$^{-1}$~\cite{tominaga2003}. They show a smaller pressure dependence (with an average pressure coefficient of 6.2~cm$^{-1}$GPa$^{-1}$ $\approx 0.79$~meV GPa$^{-1}$) in agreement with the present INS measurements and a pressure coefficient of 6.4~cm$^{-1}$GPa$^{-1}$ measured in~\cite{mita2006}. The ion displacements for both modes with irreducible representation (irrep) $A_1$ are shown in~Fig.~\ref{fig:bending} using XCrySDen~\cite{kokalj1999}. Note that in fact both modes have an out-of-plane component of the H-ion displacement, which is however larger for mode~$\gamma_{\text{OH}}$.

In our calculation, the OH-stretching mode frequencies are in the range of $2230 - 2440$~cm$^{-1}$ ($280 - 300$~meV) depending on their symmetry and propagation vector, which is close to the values calculated in~\cite{engel2018, jiang2020}. These modes are shown in Fig.~\ref{fig:stretching}. Mode $A_1$ corresponds to the hydrogen motion in the ferroelectric soft mode, which changes the direction of macroscopic polarization along the $z$ axis. Note that OH-stretching modes $A_1$, $B_1$, and $B_2$  (Fig.~\ref{fig:stretching}) preserve the value of the proton charge on the phosphate tetrahedron. Mode $A_2$ tends to increase this charge by 2 proton charges, creating a very high energy defect due to proton interaction~\cite{silsbee1964, vaks1973, abalmassov2013, beutier2015}, and as a consequence it has a higher frequency. The motion of protons in modes $B_1$ and $B_2$ creates a strong electric dipole along the $y$ and $x$ axes, respectively, which leads to a significant increase in the longitudinal mode frequency $\nu_{\text{LO}}$ for wave vectors along these axes (see Table~\ref{table:freq}). Interestingly, these frequencies almost coincide with the frequency of the $A_2$ mode, possibly because the increase in the proton interaction energy at small proton displacements in the $A_2$ mode is proportional to these displacements, and hence to the associated with them electric dipoles, as in the $B_1$ and $B_2$ modes. The frequencies of the OH-stretching modes soften with pressure (see Table~\ref{table:freq}).

We used the same approach to calculate the phonon spectrum of the FE water ice XI, in which OH-stretching modes are known to be in the range of $3100 - 3400$~cm$^{-1}$~\cite{hirsch2004, shigenari2012, zhang2016}. This benchmark  showed that our calculations underestimate these frequencies by about 150~cm$^{-1}$ (20~meV), which may be due to the PBE functional at our disposal in QE compared to calculations with RPBE~\cite{zhang2016}, PBE0~\cite{burnham2011}, meta-GGA functional~\cite{xu2019}, where agreement with experimental data is better.
Taking this into account, we could expect that the true values of the OH-stretching mode frequencies in KDP should be in the range $2380 - 2590$~cm$^{-1}$ ($295 - 320$~meV) close to experimental data~\cite{hill1968, tominaga2003}. At the same time, similar to KDP, our calculation shows a hardening of the OH-bending modes and a softening of the OH-stretching modes with pressure in ice XI. Note that this behavior was observed by Raman spectroscopy in ice Ih~\cite{minceva1984}.

\begin{figure}[t]
\centering
\includegraphics[width= \columnwidth]{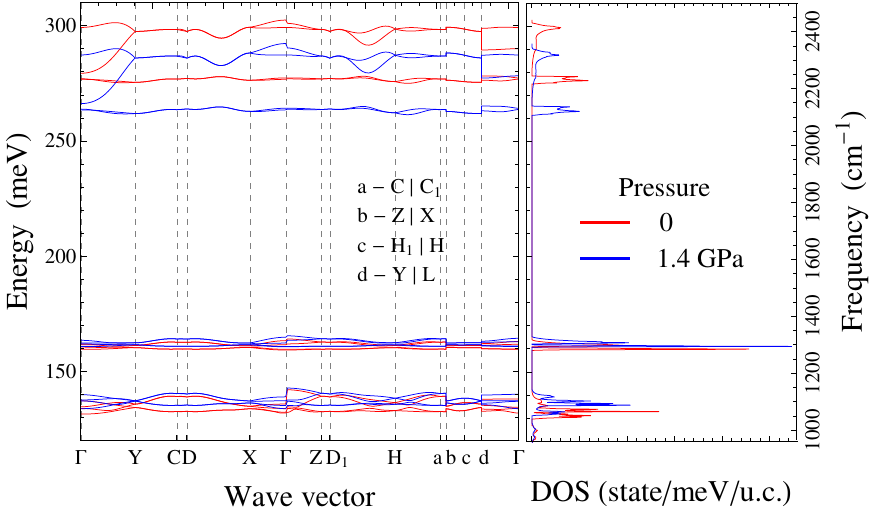}
\vspace{-0.cm}
\caption{Calculated {\it ab initio} phonon dispersion and DOS of OH-modes in KDP at zero pressure (red) and 1.4~GPa (blue). Capital letters are high symmetry points of the Brillouin zone (BZ), and lowercase letters stand for the end point of one path and the starting point of another path in the BZ, as described in the inset. The path in the BZ corresponds to that reported for KDP~\cite{menchon2018} and face-centered orthorhombic lattice~\cite{setyawan2010}. }
\label{fig:DOS}
\end{figure}

We show in Fig.~\ref{fig:DOS} for two pressures the calculated dispersion of OH-modes in KDP and their density of states (DOS), which is effectively measured in INS (contrary to the IR and Raman measurements which deal with the phonons at the $\Gamma$-point of the BZ). The result at zero pressure coincides with~\cite{menchon2018, hao2023}.  The opposite dependence on the pressure of OH-stretching and bending vibrations in the DOS is obvious.

\section{Inelastic neutron scattering}\label{INS}
\label{sec:INS}

\begin{figure}[]
\centering
\includegraphics[width= 1.0\columnwidth]{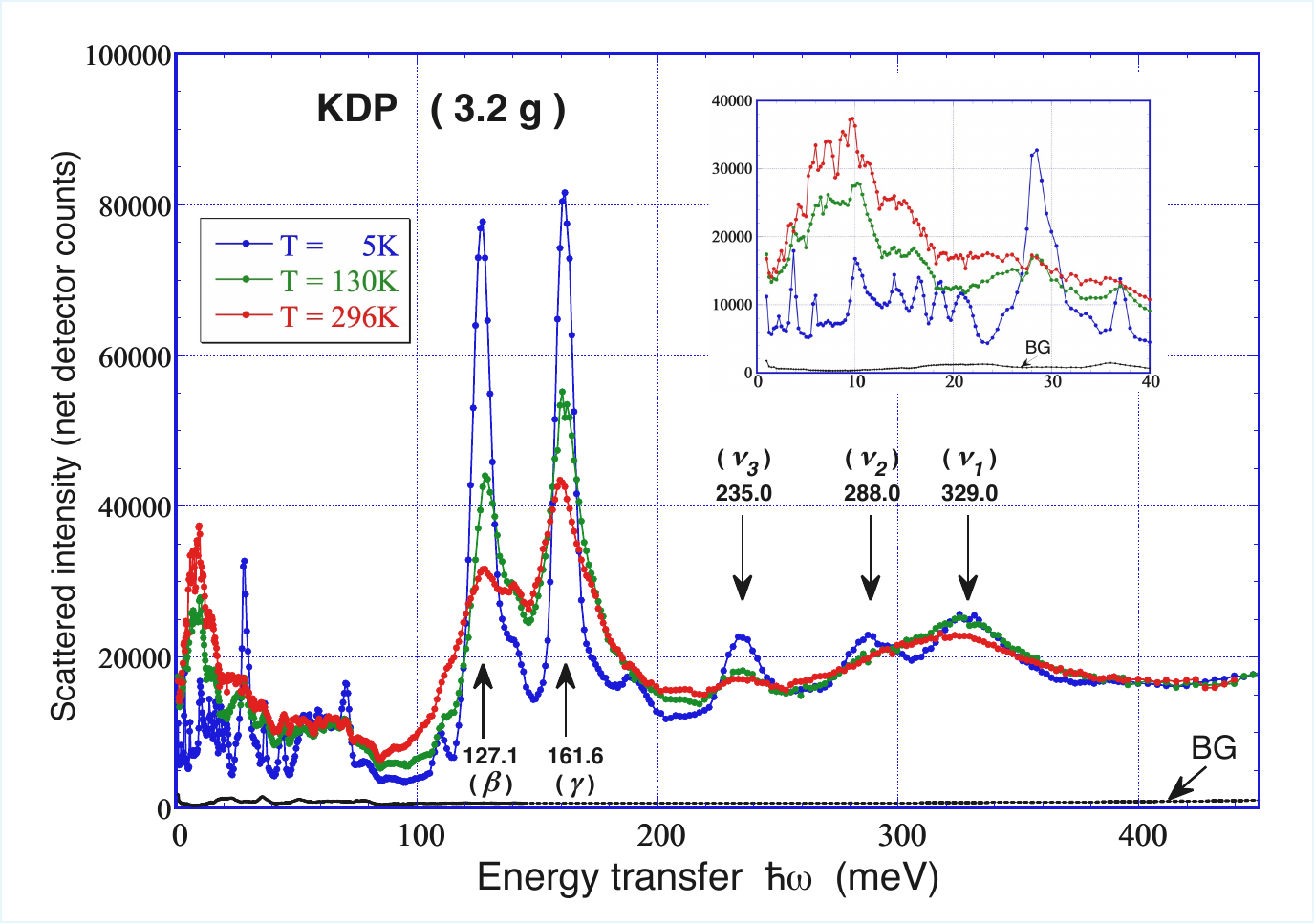}
\vspace{-0.3cm}
\caption{Inelastic neutron scattering in KDP at ambient pressure and different temperatures $T$. BG indicates experimental background. Scattered intensity is given as ``net detector counts'' collected in the `constant monitor` mode of operation (see Appendix).}
\label{fig:exp_temp}
\end{figure}

Inelastic neutron scattering measurements have been performed at the indirect geometry neutron scattering spectrometer IN1-Lagrange~\cite{ivanov2014} installed at the hot-neutron source of the high-flux nuclear reactor of the Institut Laue-Langevin, (Grenoble, France). The details of the experiment are described in Appendix.

First, in Fig.~\ref{fig:exp_temp} we present the whole energy range of atomic vibrations in KDP measured without pressure cell on a relatively big sample (total mass of 3.2~g) at three temperatures: ambient, low and near to the ferroelectric phase transition (at $T_c = 122$~K). The insert in the right upper corner outlines the low energy part of the vibration spectrum dominated by translation and libration degrees of freedom. The experimental background (cryostat together with sample holder) is practically negligible in this measurement.

The most rich spectral shapes, as expected, are observed at the base temperature (5~K). The clear sharp peaks in all parts of the spectrum progressively broaden and are partially smeared out while going to higher temperatures. Due to the fact that the neutron scattering cross section on hydrogen atoms is about an order of magnitude higher than on potassium and phosphorus the observed intensity is practically fully given by vibrations of protons in the compound, especially in the spectral range above 100~meV (800~cm$^{-1}$). There we notice the two principal characteristic hydrogen vibration lines at 127.1 and 161.6~meV (1025 and 1303~cm$^{-1}$) accompanied by three broader lines in the spectral range above 200~meV (in particular, at 235.0, 288.0 and 329.0~meV or 1896, 2323 and 2655~cm$^{-1}$).

\begin{figure}[t]
\centering
\includegraphics[width= 0.97\columnwidth]{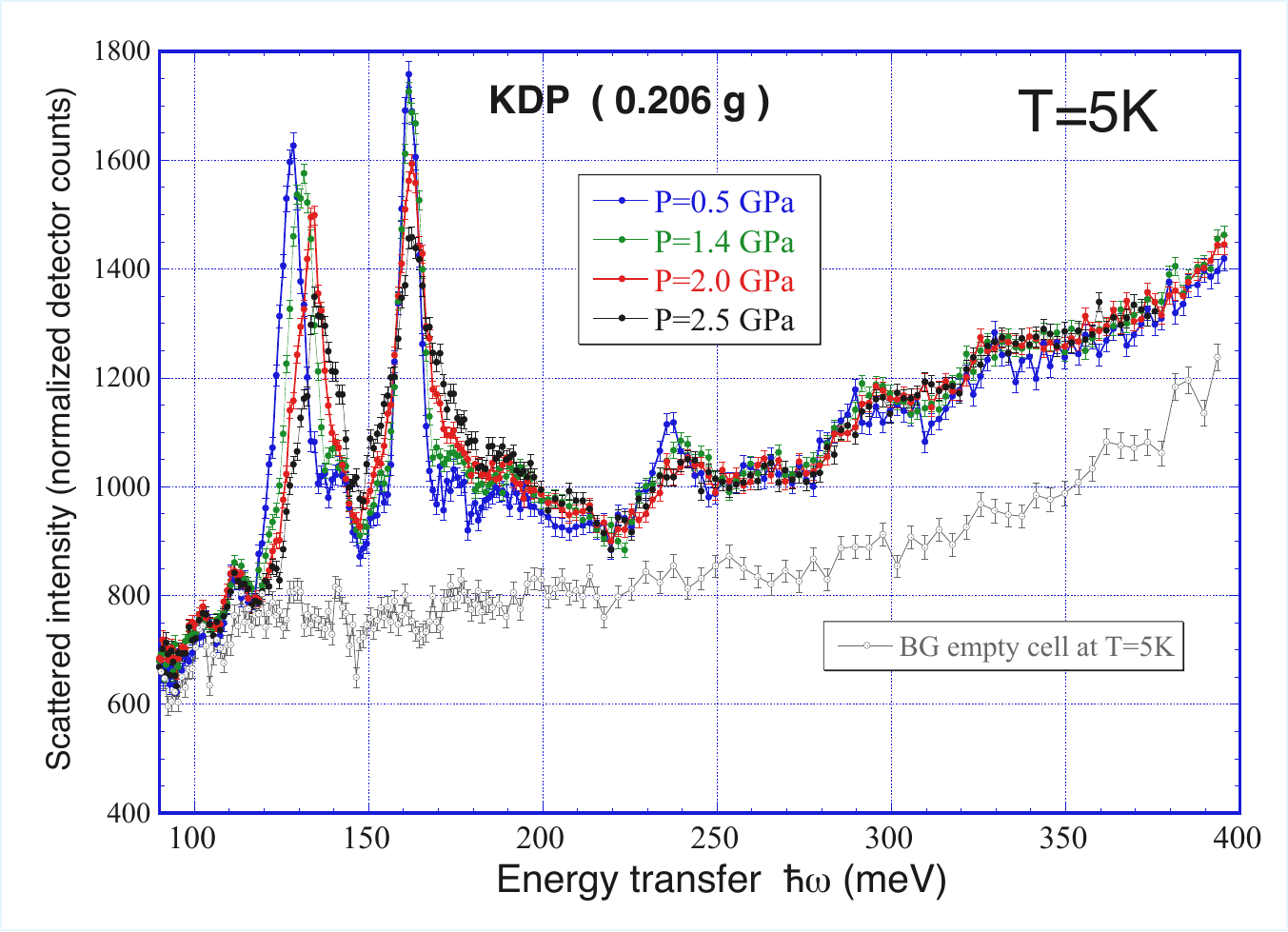}
\vspace{0.09cm}
\caption{Inelastic neutron scattering in KDP in the ferroelectric phase under different pressures $P$ at temperature $T = 5$~K. BG indicates experimental background. Measured scattered intensity is normalized to the same `constant monitor` value as in Fig.~\ref{fig:exp_temp}.}
\label{fig:exp_pressure}
\end{figure}

We also see that the signal measured in this range notably prevails the experimental background (indicated as BG in Fig.~\ref{fig:exp_temp} and Fig.~\ref{fig:exp_pressure}) manifesting effects of multiphonon neutron scattering that result, as a rule, in a relatively structureless spectral weight. However, sharp features in the density of vibration states may give rise to relatively peaked structure of the two-phonon scattering that can superpose on the spectral lines corresponding to the principal molecular vibrations of protons. So that care should be taken in assigning observed spectral lines with particular vibration frequencies. Here, once we do not have sufficient experimental means to distinguish between single vibrations and effects of combination of principal vibrations in neutron scattering (called frequently overtones), we should rely on the calculations of the expected frequencies and their behavior as a function of pressure and temperature.

In the measurements with a small sample (mass of 0.206~g) at high pressures the relatively massive pressure cell produces considerable background, in particular at lower energy range. However, in the most interesting for us high-energy vibrations range the experimental background has no remarkable structure as it is illustrated by the data obtained at the base temperature at four nominal pressures (Fig.~\ref{fig:exp_pressure}). We notice that the main spectral lines at 127.1 and 161.6~meV (1025 and 1303~cm$^{-1}$) show visible energy broadening and a shift to higher vibration energies. The pressure effect at the higher energy range is less pronounced, however, the accuracy in the peak positions is getting worse due to the lower statistical accuracy of the high-pressure data. The temperature effect at all measured pressures is similar to the observations revealed in Fig.~\ref{fig:exp_temp} at ambient pressure -- the spectral lines broaden going to higher temperatures with progressive decrease in intensity. We then collect in Table~\ref{table:freqExp} the extracted positions of the spectral peaks only at the base temperature (the phonon triplet in the range of OH-stretching modes is numbered in order from the highest to the lowest peak). There is a similar tendency for the higher energy features while the effect is less clear, partly due to the lower statistical accuracy of the high-pressure data. 

\begin{table}[b]
\setlength{\tabcolsep}{3pt}
\caption{Energies (in meV) of OH-modes in KDP from INS at different pressures~(in GPa).}
\label{table:freqExp}

\centering
\begin{tabularx}{\columnwidth}{ c | X X X X X  }
\hline 
\hline

\backslashbox{OH-mode}{Pressure} & 0 & 0.5  & 1.4 & 2.0  & 2.5   \\ \hline

$\gamma$ & 127.1 & 127.9 & 130.9 & 133.4 & 135.7 \\   
$\beta$ & 161.6 & 161.8 & 162.3 & 162.5 & 163.2 \\  
$\nu_3$ & 235.0 & 236.7 & 239.8 & 240.2 & 238.8 \\   
$\nu_2$ & 288.0 & 290.0 & 293.5 & 296.3 & 296.2 \\ 
$\nu_1$ & 329.0 & 331.5 & 331.0 & 330.4 & 331.6 \\   

\hline 
\hline
\end{tabularx}
\end{table}

\section{Discussion}\label{discussion}

Our {\it ab initio} calculation predicts a hardening of OH-bending vibrations and a softening of OH-stretching vibrations with increasing pressure in KDP, see Table~\ref{table:freq} and Fig.~\ref{fig:DOS} (we also predict similar behavior in water ice XI~\cite{hirsch2004, shigenari2012, zhang2016}, which is the low-temperature ferroelectric phase of ice Ih, where a softening of OH-stretching vibrations with increasing pressure was indeed observed using Raman spectroscopy~\cite{minceva1984}). This can be understood from the different behavior of the corresponding single-well and double-well potentials for the hydrogen atom with pressure. The single-well type potential becomes steeper when pressure is applied, whereas in the double-well type potential the barrier becomes smaller, resulting in a flatter potential near the two minima (Fig.~\ref{fig:energy_pressure}).

At the same time, our INS measurements rather indicate a hardening of phonon excitations in the region of OH-stretching modes with pressure in KDP (Table~\ref{table:freqExp} and Fig.~\ref{fig:exp_pressure}). Note that a similar situation apparently occurs in water ice, where the phonon line in the region of OH-stretching modes hardens in ice Ih with pressure~\cite{straessle2004} and for ice phases corresponding to higher pressures~\cite{li1992, li1996}.

\begin{figure}[t]
\centering
\includegraphics[width= \columnwidth]{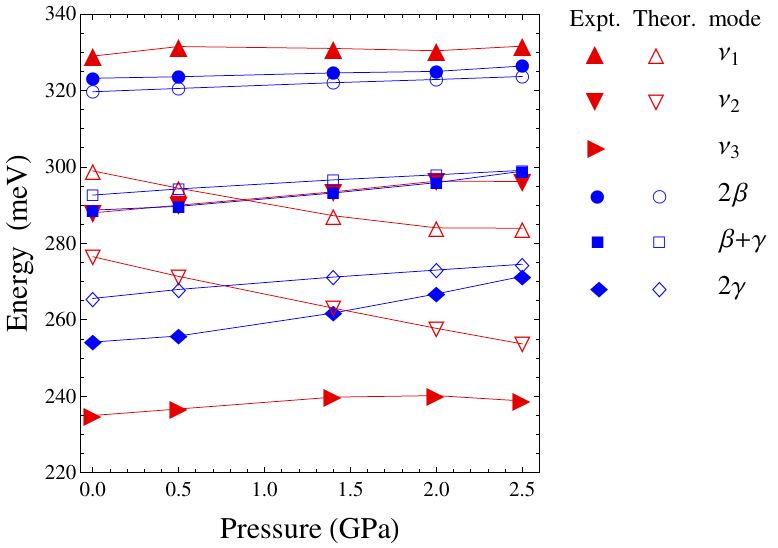}
\vspace{-0.cm}
\caption{Pressure dependence of OH-mode maxima in the region of OH-stretching modes from INS (red filled markers) and OH-stretching mode DOS maxima from {\it ab initio} calculation (the same red markers but empty). The latter are expected to be underestimated in calculations by about 20~meV. The corresponding sums of the OH-bending modes are shown in blue filled and empty markers. Mode $\nu_3$, which is not observed in calculations, may presumably be the overtone $2\gamma$, shifted down due to anharmonicity. The lines joining markers are a guide to the eye. }
\label{fig:max_pressure}
\end{figure}

Positions of the maximum values of the calculated DOS for OH-stretching modes (we take two its well distinguished maxima) and sums of OH-bending modes at different pressures are drawn in Fig.~\ref{fig:max_pressure} together with the corresponding positions from INS. We see good agreement between calculated and measured OH-bending modes with regard to their position and pressure dependence. The position and pressure dependence of the two measured  highest energies modes (labeled $\nu_1$ and $\nu_2$) in the region of OH-stretching modes almost coincide with the sums of OH-bending modes, $2 \beta$ and $\beta + \gamma$. Therefore, it can be assumed that the measured $\nu_1$ and $\nu_2$ are the overtone $2 \beta$ and the combination mode $\beta + \gamma$, respectively. Although the measured frequency of the lowest in the triplet mode $\nu_3$ is lower than twice the frequency of the OH-bending mode $2 \gamma$, this may be due to the anharmonism of the bending mode, and $\nu_3$ is actually an overtone of this mode. Strong anharmonicity in phonon spectra can at present be calculated from first principles with some additional effort~\cite{ribeiro2018, romero2015, subedi2015}, but we do not expect this to change the pressure dependence of the OH-modes considered here, so a detailed study of this issue is beyond the scope of this paper.

Below we present some arguments why the OH-stretching modes are not observed in our INS experiment. First, we note that the neutron momentum transfer in our experiment is large, being $Q^2 = \hbar \omega / 2.0717$~\AA$^{-2}$/meV, where $\hbar \omega$ is the energy transfer. Given the almost flat dispersion of OH-modes, we neglect the interaction between protons and consider incoherent neutron scattering on a single proton. The scattering cross section is proportional to the modulus square of the matrix element between the proton ground state $| 0 \rangle$ and its excited state $| n \rangle$ in a potential created by the oxygen atom, $\sigma \propto |\langle n | \exp(-i {\bf Q} {\hat {\bf u}}) | 0 \rangle|^2$, where ${\bf Q}$ refers to the neutron momentum transfer and ${\hat {\bf u}}$ is the proton coordinate operator. For the states near the ground state, this potential is almost harmonic and the scalar product can be rewritten as a sum over projections onto the principal axes of this potential, ${\bf Q} {\hat {\bf u}} = \sum_{\alpha} Q_{\alpha} {\hat u}_{\alpha}$. In the harmonic approximation, the proton coordinate can be represented through the ladder (creation and annihilation) operators as ${\hat u}_{\alpha} = u_{\alpha} ({\hat a}_{\alpha} + {\hat a}_{\alpha}^{\dag})$, where $u_{\alpha} = (\hbar/2 m\omega_{\alpha})^{1/2}$ is the root mean square displacement of the proton in the ground state. Expanding the exponential up to quadratic terms, we get $\sigma \propto |\langle n | i\sum_{\alpha} Q_{\alpha} {\hat u}_{\alpha} + (1/2)\sum_{\alpha} Q_{\alpha}^2 {\hat u}_{\alpha}^2 + \sum_{\alpha \neq \beta} Q_{\alpha} Q_{\beta} {\hat u}_{\alpha} {\hat u}_{\beta}| 0 \rangle|^2$, where the only nonzero matrix elements are $\langle 1 | {\hat a}^{\dag}_{\alpha} | 0 \rangle = 1$ and $\langle 2 | {\hat a}^{\dag}_{\alpha} {\hat a}^{\dag}_{\alpha} | 0 \rangle = \sqrt{2}$. The first term of the sum is responsible for the fundamental phonon lines in neutron scattering, and the other two correspond to their overtones and combinations, respectively. Thus, the cross section is proportional to $(Q_{\alpha} u_{\alpha})^2$ for fundamentals, $(1/2)(Q_{\alpha} u_{\alpha})^4$ for overtones, and $(Q_{\alpha} Q_{\beta} u_{\alpha}u_{\beta})^2$ for combinations. For OH-mode fundamental frequencies $\nu_{\gamma} \approx 130$~meV, $\nu_{\beta} \approx 160$~meV, and $\nu_{\nu} \approx 280$~meV~(from Table~\ref{table:freq}) we get $u_{\gamma} \approx 0.126$~\AA, $u_{\beta} \approx 0.113$~\AA, and $u_{\nu} \approx 0.086$~\AA\, and $Q_{\gamma} \approx 7.92$~\AA$^{-1}$, $Q_{\beta} \approx 8.79$~\AA$^{-1}$, and $Q_{\nu} \approx 11.63$~\AA$^{-1}$ with a constant value of the product $Q_{\alpha} u_{\alpha} \approx 1$ for fundamentals. However, for overtones and combinations of OH-bending modes, the square of the momentum transfer $Q^2$ is approximately twice as large as for fundamentals (according to its dependence on the energy transfer $\omega$ in our experiment), and, consequently, $Q_{\alpha} u_{\alpha} \approx 2$. This leads to an increase in the scattering cross section for overtones by approximately 2 times, and for combinations by 4 times compared to the fundamentals.

Secondly, INS scattering is also affected by the Debye-Waller (DW) factor, $\exp(-2W) = \exp(-\langle({\bf Q} {\bf u})^2\rangle)$~\cite{furrer2009}, which is determined by the projection of the proton displacement ${\bf u}$ on the neutron momentum transfer ${\bf Q}$, which is perpendicular for bending and stretching vibrations.  The total proton displacement includes the contribution from all phonon modes~\cite{landaulifshitz5}, and can be larger along the OH-bond, thereby diminishing the DW factor for OH-stretching modes. In other words, the along-bond motion of the proton can be
considered as having three principal components: the along-bond motion of the oxygen atom, to which the proton is bound, the variation in the instantaneous OH distance, and the along-bond motion of proton in the static potential~\cite{nelmes1987a}, which is given by $u_{\nu}$ at temperatures lower than the vibration frequency. As a result, the probability distribution function of protons turns out to be almost isotropic with the measured thermal parameter along the H-bond $u_{\nu}^{\text{expt}} \approx 0.13$~\AA\, at $T = T_c -20$~K~\cite{nelmes1987}. Thus, the DW factor for the OH-stretching fundamental modes is approximately the same as for the overtones and combinations of OH-bending modes, $\exp(-2W_{\nu}) \approx 0.07$ with $\omega_{\nu} \approx 320$~meV and  $u_{\nu}^{\text{expt}}$, and their relative intensities are determined by the square of the matrix element, as discussed above. At the same time, for OH-bending mode fundamentals the DW factor is larger, $\exp(-2W_{\gamma, \beta}) \approx 0.37$ with $\omega_{\gamma, \beta}$ from Table~\ref{table:freqExp} and $u_{\gamma, \beta}$ calculated above, due to the smaller neutron momentum transfer $Q$. This is consistent with the line intensities in Fig.~\ref{fig:exp_temp} and Fig.~\ref{fig:exp_pressure}. We note that it was earlier shown that multiphonon neutron scattering in hydrogen-bonded materials can have intensity comparable to the one-phonon scattering~\cite{springer1978, mestnik1994, kolesnikov1997, cheng2019}.

Although KDP becomes a quantum paraelectric at pressures above $p_c \approx 1.7$~GPa~(at zero temperature)~\cite{samara1971, samara1987}, which is thought to be due to proton tunneling in the double-well potential along the hydrogen bond~\cite{blinc1966b, abalmassov2013b, abalmassov2016} (the latter is questioned in some works~\cite{tokunaga1987, ichikawa1987, mcmahon1990, sugimoto1991, ikeda1998, endo2002PRL, merunka2007} though), we can not see this quantum phase transition in our {\it ab initio} calculation since all nuclei, including protons, are treated as classical. On the other hand, the OH-modes $\nu_1 - \nu_3$ in the range 200 -- 350~meV from INS seem to be unaffected by this transition~(Fig.~\ref{fig:exp_pressure}), whereas their shape changes significantly at the ferroelectric phase transition at $T_c = 122$~K at atmospheric pressure~(Fig.~\ref{fig:exp_temp}). At the same time, the width of the OH-bending modes appears to narrow at pressures around $p_c$~(Fig.~\ref{fig:exp_pressure}), which will be studied further in the future.

To further clarify the question of the identity of OH-modes observed in INS in the region of OH-stretching modes, it is necessary to measure the dependence of the cross section on the neutron momentum transfer $Q$, which is possible at some installations~\cite{kolesnikov2016, hattori2022}. Raman spectroscopy, in which the overtones are strongly suppressed due to the small photon momentum transfer $Q$, can also help clarify the situation.

\section{Conclusion}\label{conclusions}

In this study, we aimed to determine the nature of the phonon triplet in the range of OH-stretching modes in KDP observed experimentally. While our {\it ab initio} calculations predicted hardening of OH-bending modes and softening of OH-stretching modes, all OH-modes observed in our INS experiment hardened with pressure. This suggests that the phonon triplet in the range of OH-stretching modes in KDP observed in our INS is mainly due to combinations and overtones of OH-bending modes, the scattering cross section of which we estimate to be at least twice as large in our experiment as for the fundamental OH-stretching modes. The results obtained may also be applicable to INS experiments with other hydrogen-bonded materials.


\begin{acknowledgments}
The authors thank Monica Jimenez-Ruiz (ILL, France) for her help and assistance during the neutron scattering measurements. The work of V.A.A. was carried out within the framework of the state contract of the Sobolev Institute of Mathematics, Project No. FWNF-2022-0021. The Siberian Branch of the Russian Academy of Sciences (SB RAS) Siberian Supercomputer Center is gratefully acknowledged for providing supercomputer facilities.
\end{acknowledgments}

\appendix*

\section{Details of INS measurements}
\label{sec:INS_details}

The scattered neutron energy was fixed at 4.5~meV or 36.3~cm$^{-1}$ with pyrolytic graphite analyser. We used different double-focussing monochromators to vary the energy of incident neutrons. The low energy part of the vibration spectra was measured with elastically bent silicon crystals (incident energy up to 16~meV or 129~cm$^{-1}$ with crystal reflection Si111 and up to 30~meV or 242~cm$^{-1}$ with Si311) while for the higher energy part the mosaic crystals of copper were chosen (up to maximum energy of 500~meV or 4000~cm$^{-1}$ with crystal reflection Cu220). The energy resolution in different spectral ranges was about $1.8 - 2.5$\% of the vibration energy given by the difference of the incident and scattered neutron energies. All neutron scattering measurements have been performed at fixed number of monitor counter installed in the incident neutron beam. This mode of operation together with the chosen monitor efficiency to be inverse proportional to the incident neutron velocity allows the counted scattered intensity to be used as the density of phonon states in the incoherent approximation.

The used neutron spectrometer IN1-Lagrange works with a relatively small fixed final energy $E_f = 4.5$~meV (corresponding to the neutron wave vector $k_f = 1.47$~\AA$^{-1}$) compared to the measured energy of proton vibrations (several hundred meV). This results in a situation when the dynamic range accessible at this spectrometer, or the effectively probed region in the momentum transfer - energy transfer plane ($Q$, $\hbar \omega$) is given by a relatively narrow band, approximately of width $k_f$, stretched around the line $\hbar \omega = (\hbar^2/2m_n) (k_i^2-k_f^2) \approx (\hbar^2/2m_n) k_i^2 \approx \hbar^2 Q^2/2m_p$, where $m_n$ is the neutron mass, $m_p$ is the the proton mass, $\hbar$ is the reduced Planck constant. Consequently, this condition is approximately valid for all the analyzed spectral lines, because for each energy the integration of the scattered intensity is done in a narrow range around the corresponding $Q$.

The samples of KDP for neutron scattering experiments were prepared from commercially available powder material. For the measurements without pressure cell the chosen amount of sample was wrapped in a piece of thin aluminium foil. The pressure cell for this experiment was manufactured in the clamp cell scheme\cite{klotz2012} ``cylinder + belt'' with Ni-Cr-Al alloy as cylinder material and Ti-metal as belt material. The inner hollow cylinder had a sample volume of 4~mm in diameter and about 10~mm in length. The ``lower'' end of the cell was mechanically rounded in order to respect the open geometry for scattered neutrons on the IN1-Lagrange spectrometer. A corresponding sample powder share (weighed about 200~mg) was inserted into a thin polyethylene cylinder together with the pressure transmitting medium (Fluorinert liquid) and closed with a polyethylene cork. The static pressure was applied using a standard press with help of a tungsten carbide piston acting on the cork through a friction reducing bronze ring. The nominal pressure was defined as the applied force divided by the surface of the sample channel cross section. The piston position after applying pressure was secured by a special nut so that the pressure cell could be then transported to the neutron spectrometer. 

The prepared sample in the pressure cell was inserted into a dedicated closed cycle refrigerator on the neutron spectrometer where it was cooled to a chosen temperature. The temperature was maintained with a precision of about 0.1~K during neutron scattering measurements. We did not control possible pressure variation on cooling and the reported pressure values correspond to nominal pressure applied at ambient temperature conditions.

\bibliography{kdp_neutrons_v16.bbl}

\end{document}